\documentstyle[12pt]{article}
\begin{document}
\bibliographystyle{unsrt}
\vbox {\vspace{6mm}}

\begin{center}
{\bf PURIFICATION OF IMPURE DENSITY OPERATORS AND THE RECOVERY
OF ENTANGLEMENTS}
\end{center}

\begin{center}
V. I. Man'ko\footnote{On leave from the Lebedev Physical
Institute, Moscow, Russia.}, G. Marmo,
E. C. G. Sudarshan\footnote{Permanent address:Physics
Department, Center for Particle Physics, University of Texas, Austin,
Texas, 78712 USA.} and F. Zaccaria
\end{center}

\begin{center}
 Dipartimento di Scienze Fisiche\\
Universit\'a di Napoli "Federico II" and\\
Istituto Nazionale di Fisica Nucleare, Sezione di Napoli\\
Mostra d'Oltremare, Pad.~20 -- 80125 Napoli, Italy
\end{center}

\begin{abstract}
The need to retain the relative phases in quantum mechanics implies
an addition law parametrized by a phase of two density operators required
for the
purification of a density matrix. This is shown with quantum tomography and the
Wigner function. Entanglement is determined in terms of phase dependent
multiplication.
\end{abstract}


\noindent

Quantum mechanics as traditionally formulated[1]
involves three principles. The states of the system are normalized
vectors in a Hilbert space, (selfadjoint) linear operators correspond
to (real) dynamical variables; and the expectation value for any dynamical
variable is bilinear (sesquilinear) in the state vectors. But
the overall phase of the state vector is irrelevant in those
computations involving only that one state. The state corresponds to a
``ray'' in Hilbert space
$\left\{\psi : e^{i \alpha}
\psi _0\right\}.$
Alternatively
the expectation value can be expressed by means of  the density matrix
$$\rho=\psi\psi^\dagger=\psi_0\psi_0^\dagger,$$
which is defined on the ray.

These density operators are ``pure'' and of rank one:
$$
\rho^\dagger=\rho\,;\qquad \rho>0\,;\qquad \mbox{tr}\,\rho=1\,;
\qquad \rho^2=\rho\,.$$
There is a probability addition of the density operators.
If $\rho_1$ and $\rho_2$ are two pure density matrices
$$
\rho=\cos^2\Theta\rho_1+\sin^2\Theta\rho_2
$$
also is a density matrix; but it is $\underline{\mbox{not~pure}}$,
but mixed.  These are the interior points of the convex set of density
operators.  Hence

So
$$
\rho^2\neq\rho\,,\qquad \rho-\rho^2>0\,.
$$
Can we have an additional operation of constructing a {\bf pure}
density operator from the {\bf sum} of the two density operators
$\rho_1,~\rho_2?$
We must have such a construction since the superposition principle
of quantum mechanics[1] tells us how to add two state
vectors $\psi_1$ and $\psi_2$ to form a superposed state vector
$$
\psi=\left\{\cos \Theta~\psi_1+e^{i\varphi^1}\sin\Theta~\psi_2\right\}
$$
which is normalized if $\psi_1,~\psi_2$ are orthonormal.
For nonorthogonal states, we have to use the normalization factor
$$
\left(1+\sin 2\Theta~\cos\varphi\mid\langle \psi_1\mid
\psi_2\rangle\mid\right)^{-1/2}.
$$
where $\varphi$ includes the phase of $\langle \psi_1\mid
\psi_2\rangle$ along with $\varphi^1$.  In terms of density operators,
the
$\varphi$-addition law which we introduce is
$$\rho (x,y, \varphi)=
\{\cos \Theta\psi_1(x)+e^{i\varphi}\sin\Theta\psi_2(x)\}
\{\cos \Theta\psi_1^\dagger(y)+e^{i\varphi}
\sin\Theta\psi_2^\dagger(y)\}$$
$$= \cos^2\Theta \rho_1(x,y)+\sin^2\Theta \rho_2(x,y)+
\sin 2\Theta\ cos\varphi \rho_{12}(x,y,\varphi).$$

In terms of the Wigner function[2],
we get a $\varphi$-addition law which
has an interference term proportional to $\sin 2\Theta\cos \varphi.$, i.e.
$$
W\left(q,p\right)=\cos^2\Theta ~W_1\left(q,p\right)
+\sin^2\Theta ~W_2\left(q,p\right)+\sin 2\Theta~ \cos\varphi~
I_{12}\left(q,p, \varphi\right)
$$
where $I_{12}$ is the (generalized) Wigner function corresponding to
the  operator having the structure of the root square of a convolution
of the product of the two density matrices, which we symbolically denote as
$$
I_{12}\rightarrow \sqrt{\rho_1\rho_2}\,.
$$
There is thus a one-parameter addition law of density operators and of Wigner
functions
with probabilities $\cos^2\Theta$ and $\sin^2\Theta $ and with extra
 intereference term.
Note that this  $W\left(q,p\right)$ is pure and satisfies the
purity criterion
$$
\int\int dp~dq\left\{W\left(q,p\right)\right\}^2=\frac {1}{2\pi}\,.
$$

For quantum tomography[3],
 also we can construct superposition of tomograms
using a one parameter addition law:
$$
\Phi\left(\lambda,\mu;x\right)=\cos^2\Theta
~\Phi_1\left(\lambda,\mu;x\right)
+\sin^2\Theta~ \Phi_2\left(\lambda,\mu;x\right)
+\sin 2\Theta ~\cos\varphi~\Phi_{12}
\left(\lambda,\mu;x\right)
.
$$

Here tomograms $\Phi_{1}.\Phi_{2}$ determine the probability density
of
 quadrature $x = \lambda q + \mu p$ in the pure states $\psi_{1}$ and
$\psi_{2}$. The tomogram $\Phi_{12}$ corresponds to the interference
term $I_{12}.$

The passage from the impure density operator
$\cos^2\Theta\rho_1+\sin^2\Theta \rho_2$
to the pure $\varphi$-dependent addition may be called
$\underline{\mbox{purification}}.$
Note that the purification introduces the relative phase $\varphi$
which was not in $\rho_1$ or $\rho_2.$

The density operator of a composite system $AB$ with subsystems
$A$ and $B$ may be chosen pure or impure. For a pure density operator
$\rho_{AB},$ one can get the density operator $\rho_A$ and $\rho_B$
by the partial trace operation
$$
\rho_A=\mbox{tr}_B\,(\rho_{AB});\qquad
\rho_B=\mbox{tr}_A\,(\rho_{AB})\,.
$$
It is not necessary that $A$ and $B$ have the same dimensionality.
Unless $\rho_{AB}$ is a direct product of pure states of $A$ and $B,$
a pure $\rho_{AB}$ yelds impure $\rho_A$ and $\rho_B.$ But they will
have the same rank $R$ and the same nonnegative eigenvalues which sum
up to unity.
The density operator
$$
\rho'_{AB}=\rho_A\otimes\rho_B\neq\rho_{AB}
$$
is impure.
Thus, the whole is greater than the parts: there is additional information
in $\rho_{AB}.$ These are the ``entanglement'' terms[4].

We purify the product
 $\rho'_{AB}$ in the same way that we used before for the mixture
of two density operators $\rho_1$ and $\rho_2.$ Here we have $n$ such
pure states mixed together and need $(n-1)$ phase angles
$\varphi_1=0,\varphi_2,\ldots\varphi_n.$
The diagonal form of $\rho'_{AB}$ has only $n$ nonzero diagonal elements.
We need to introduce the offdiagonal elements
$$
\sqrt{\lambda_j\lambda_k}\,e^{i\left(\varphi_j-\varphi_k\right)}
$$
in the $\left(j,k\right)$ location. Note that while we have $ n(n-1)/2$
offdiagonal terms, there are only $(n-1)$ phases $\varphi_j.$

The purification of the density matrix $\rho'_{AB}$ we call as
the $\varphi$- multiplication law of the density matrices $\rho_{A}$ and
$\rho_{B}.$

While purification of an impure density addition is dependent on one
phase angle, the form of the
 entanglement is constructed depending on $R-1$ phase angles. These have
to be obtained
 from other considerations.

The same kind of $\varphi$-addition law and $\varphi$-multiplication
law holds for other representatives like the quantum tomograms, the
diagonal coherent state distribution function in quantum optics[5] and
the Husimi - Kano[6] density of coherent state projection
 operators. We expect to return to this discussion elsewhere.

\section*{Acknowledgements}

\noindent

V.I.M. thanks Dipartimento di Scienze Fisiche
Universit\'a di Napoli ``Federico II'' for kind hospitality
and Russian Foundation for Basic Research for partial support.

\end{document}